\documentclass[final]{pasj00}
\SetRunningHead{T. Akahori and K. Masai}{Core Structure of Intracluster Gas}
\title{Core Structure of Intracluster Gas: Isothermal Hydrostatic Equilibrium}
\author{Takuya \textsc{Akahori} and Kuniaki \textsc{Masai}}
\affil{
Department of physics, Tokyo Metropolitan University, Hachioji, Tokyo 192-0397}
\email{akataku@phys.metro-u.ac.jp, masai@phys.metro-u.ac.jp}
\Received{2004/07/26}
\Accepted{$\langle$acception date$\rangle$}
\Published{$\langle$publication date$\rangle$}

\begin{document}

\maketitle

\begin{abstract}

We investigate core structures of X-ray emitting intracluster gas based on the so-called $\beta$-model, which is an isothermal hydrostatic model often used in observational studies. 
We reconsider the $\beta$-model and find that the virial temperature $T_{\rm vir}$ of a cluster may be represented better by $\beta T_X$ than $T_X$, where $\beta$ is the parameter obtained from the X-ray surface brightness and $T_X$ is the emission-weighted mean temperature of the gas. We investigate 121 clusters observed by {\it ROSAT} and {\it ASCA} and find that the luminosity-temperature relation $L_X - \beta T_X$ is less steep than $L_X - T_X$.
We classify the clusters into two core-size groups in order to investigate their properties in detail.  While in the larger core group the core radius is marginally proportional to the virial radius, no significant relation is found for the smaller core group.  This may suggest that the smaller cores reflect the presence of cD galaxies, effect of radiative cooling or asymmetry in the surface brightness.  We examine such possibilities, and find that the clusters of smaller cores have shorter cooling time than the Hubble time, while no significant correlation is found with cD or asymmetry.
We carry out hydrodynamical calculations to simulate the $\beta$-model, intending to see the behavior of the isothermal gas under the gravitational potential including the dark matter and galaxies with or without a central cD galaxy.  Calculations show $r_c\propto r_{\rm vir}$ and $T_{\rm vir}\simeq\beta T$ consistently with our consideration to the $\beta$-model.
Also is found from calculations that the presence of a large cD galaxy may form a gas core $\sim 40$~kpc, which seems too small to account for the range of the core sizes, 40--80~kpc, of the smaller core group.

\end{abstract}

\section{Introduction}

Galaxy clusters are the self-gravitational systems dominated by the dark matter.
The overall structure of the intracluster gas can be explained by the $\beta$-model, which consists of a flat core part and the envelope. In the $\beta$-model, the core radius and the envelope slope represented by $\beta$ are important parameters to characterize the structure of the gas. 

In simple models of structure formation by gravity, galaxy clusters constitute a homologous family (see e.g., \cite{Kaiser86}). Clusters are self-similar in shape and their properties relate in predictable scaling laws. For example, the core radius is expected to be proportional to the virial radius of the cluster.

Observations show, however, that core radii of clusters are not simply scaled.  For 79 clusters observed by {\it ROSAT} and {\it ASCA}, \citet{OM02} find two typical core sizes or two core-size groups: the core radius distribution exhibits two distinct peaks at $\sim 50$~kpc and $\sim 200$~kpc for $H_0=70~{\rm km~s^{-1}~Mpc^{-1}}$ (\cite{OM04}).  They also show that some clusters are expressed better by the double $\beta$-model with two cores than the single $\beta$-model, and then the radii of the larger and smaller cores fall into the above two core-size groups respectively.  These results can be an important clue for understanding the formation and evolution of cluster cores. 

Observations also show that the mass-temperature or luminosity-temperature relation of clusters do not simply scale self-similarity (e.g., \cite{Markevitch98}; \cite{AE99}; \cite{F01}; \cite{Sanderson03}). Cosmological hydrodynamics calculations suggest that various-sized clusters are formed through the structure formation in the universe. But such clusters are neither isothermal (e.g., \cite{Borgani04}) nor those expressed by the $\beta$-model.

As for the core structure, effects of radiative cooling can be a considerable factor. For instance, the cooling flows may increase the density of the cluster central region (see \cite{Fabian94} for a review), though recent {\it ASCA}, {\it Chandra} and {\it XMM-Newton} observations (e.g., \cite{Makishima01}; \cite{Kaastra01}; \cite{Peter01}; \cite{Lewis02}) show  several results inconsistent with the standard cooling flow model. Although various models have been proposed for thermal balance against cooling flows,  (see e.g., \cite{Boehringer02}; \cite{Peter03} references therein), quantitative account is not yet given successfully.

Another factor of the double core may be the presence of a cD galaxy in the cluster central region (e.g., \cite{JF84}; \cite{Xu98} and references therein).  However, the cores associated with cDs seem small compared to the typical core size of the smaller core group; for example, in the Centaurus cluster, the core radius due to the cD is estimated to be $\sim$ 40~kpc or less (see \cite{Ikebe99}). 

In the present paper, we investigate core structures of the isothermal intracluster gas.  The origin of the two core sizes is of our interest. In section~\ref{sec:theory}, we examine the $\beta$-model considering the isothermal X-ray temperature and the virial temperature. In section~\ref{sec:observation}, we investigate 121 clusters observed by {\it ROSAT} and {\it ASCA} and discuss the properties of the smaller and larger core clusters. In section~\ref{sec:simulation}, we show the results of hydrodynamical calculations to simulate the $\beta$-model or the behavior of the isothermal gas under the cluster potential, and discuss the effect of a central cD galaxy. We summarize the present work in section~\ref{sec:remarks}.

\section{Isothermal $\beta$-model}\label{sec:theory}

The $\beta$-model has been widely used in analyses of the X-ray emitting hot gas in clusters of galaxies. This is an isothermal hydrostatic model based on the approximate King model for galaxies in clusters (see e.g., \cite{Sarazin86}). The density profile of the intracluster gas based on the $\beta$-model is expressed as 
\begin{equation}
\rho_g(r)=\rho_{g0}[1+(r/r_c)^2]^{-3\beta/2},
\label{eq:betamodel}
\end{equation}
where $\rho_{g0}$ is the gas density at the cluster center and $r_c$ the core radius. These quantities and $\beta$ are obtained from the X-ray surface brightness profile for isothermal temperature $T_X$. The core radius $r_c$ is defined originally as a radius at which the projected galaxy density falls to half its central value (\cite{Sarazin86}), and characterizes the gas distribution as well. 

The total mass enclosed within radius $r$ is obtained as
\begin{equation}
M(r)=\frac{3kT_X\beta r}{\mu mG}\frac{(r/r_c)^2}{1+(r/r_c)^2},
\label{eq:m(r)}
\end{equation}
where $T_X$ is the gas temperature inferred from X-ray observations mentioned above, $\mu$ the mean molecular weight and $m$ is the hydrogen mass. Thus, the total mass integrated up to the virial radius $r_{\rm vir}$ of the cluster comes to
\begin{equation}
M(r_{\rm vir})=\frac{3kT_X\beta r_{\rm vir}}{\mu mG}\frac{x_{\rm vir}^2}{1+x_{\rm vir}^2},
\label{eq:mvir}
\end{equation}
where $x_{\rm vir}\equiv r_{\rm vir}/r_c$.

\begin{figure}
\begin{center}
\FigureFile(85mm,60mm){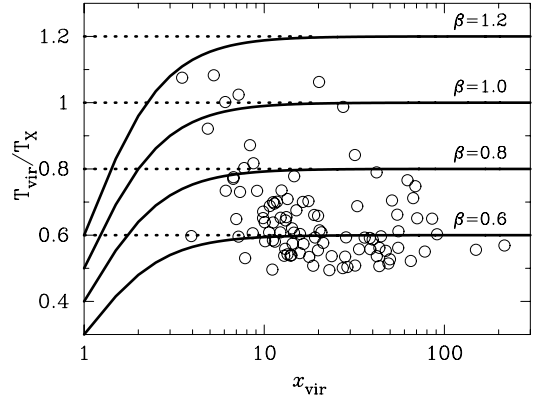}
\end{center}
\caption{
The ratio of the virial to isothermal temperatures is plotted against $x_{\rm vir}$. Each solid line represents the relation given by equation~(\ref{eq:Tvir-Tiso}) for $\beta$=0.6, 0.8, 1.0, and 1.2. The open circles represent $\beta x_{\rm vir}^2/(1+x_{\rm vir}^2)=T_{\rm vir}/T_X$ for distant clusters (\cite{OM02}; see section \ref{sec:observation}).}
\label{fig:1}
\end{figure}

On the other hand, for a virialized cluster, the virial mass is given by
\begin{equation}
M_{\rm vir}=\frac{3kT_{\rm vir}r_{\rm vir}}{\mu mG},
\label{eq:virialtheorem}
\end{equation}
where $T_{\rm vir}$ is the virial temperature. 
For $M({r_{\rm vir}})=M_{\rm vir}$, we have the relation between the two temperatures as
\begin{equation}
T_{\rm vir}=
\frac{x_{\rm vir}^2}{1+x_{\rm vir}^2} \beta T_X.
\label{eq:Tvir-Tiso}
\end{equation}
The observed values of $\beta x_{\rm vir}^2/(1+x_{\rm vir}^2)=T_{\rm vir}/T_X$ for distant clusters (\cite{OM02}; see section \ref{sec:observation}) are plotted in figure~\ref{fig:1}. 

\begin{figure}
\begin{center}
\FigureFile(85mm,60mm){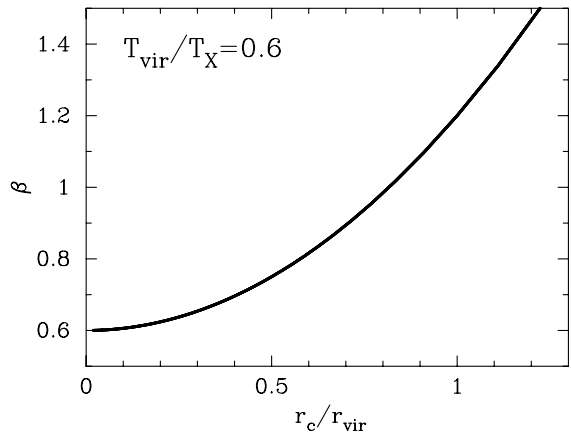}
\end{center}
\caption{Relation between $r_c/r_{\rm vir}$~$(=1/x_{\rm vir})$ and $\beta $ for $T_{\rm vir}/T_X=0.6$.}
\label{fig:2}
\end{figure}

Many of the clusters show $x_{\rm vir}\gg 1$, and therefore $T_{\rm vir}/T_X\simeq\beta\sim 0.6$ according to equation~(\ref{eq:Tvir-Tiso}).
This means that $T_X$ could deviate from $T_{\rm vir}$ depending on $\beta$, and $T_{\rm vir}$ is well represented by $\beta T_X$ rather than $T_X$ when $x_{\rm vir}\gg 1$. 
In virialized clusters, $kT_{\rm vir}/\mu m\sim\sigma _r^2$ for the line-of-sight velocity dispersion $\sigma _r$ of galaxies, so that
\begin{equation}
\beta _{\rm spec}T_X\simeq T_{\rm vir}
\end{equation}
is expected with
\begin{equation}
\beta _{\rm spec}\equiv {\sigma_{\rm gal}^2 \over \sigma_{\rm gas}^2} \simeq 
\frac{\mu m\sigma _r^2}{kT_X},
\end{equation}
where $\sigma _{\rm gal}^2$ and $\sigma _{\rm gas}^2$ are the velocity dispersions of galaxies and gas, respectively. Equation~(\ref{eq:Tvir-Tiso}) and figure~\ref{fig:1} suggest 
\begin{equation}
\beta T_X\simeq T_{\rm vir}
\label{betaTiso-Tvir}
\end{equation}
or $\beta\simeq \beta _{\rm spec}$ for $x_{\rm vir}=r_{\rm vir}/r_c \gg 1$. 
Otherwise, if $x_{\rm vir}$ is not enough large, $\beta$ would deviate from $\beta _{\rm spec}$ depending on $x_{\rm vir}$ for given $T_{\rm vir}/T_X$. 
Figure~\ref{fig:2} shows the relation between $r_c/r_{\rm vir}$~$(=1/x_{\rm vir})$ and $\beta$, where we take $T_{\rm vir}/T_X=0.6$. With increasing $r_c/r_{\rm vir}$ (or decreasing $x_{\rm vir}$), $\beta$ increases monotonically. Actually, such tendency is found by observations (e.g., \cite{NA99}). 

\begin{figure*}
\begin{center}
\FigureFile(80mm,60mm){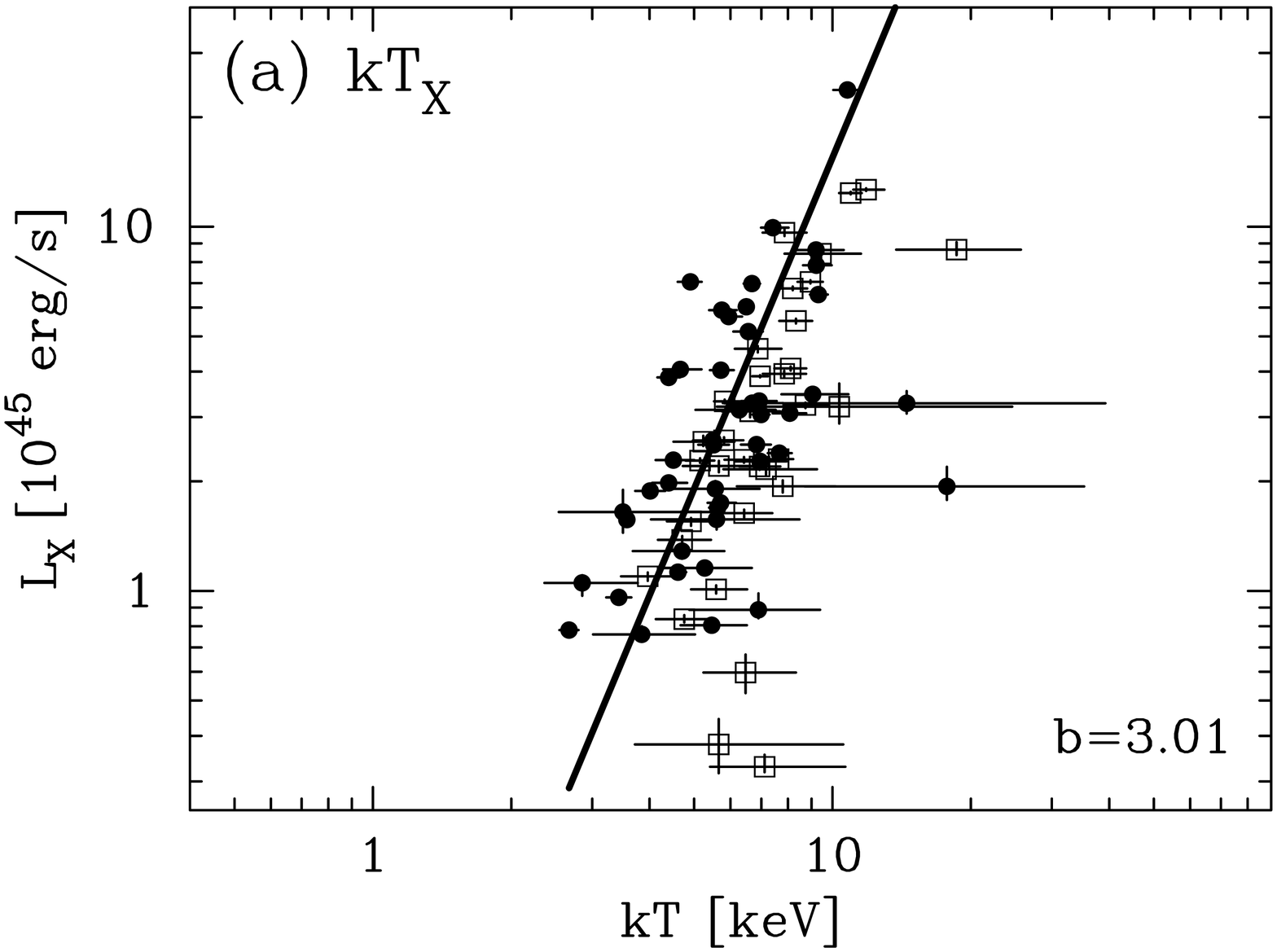}
\FigureFile(80mm,60mm){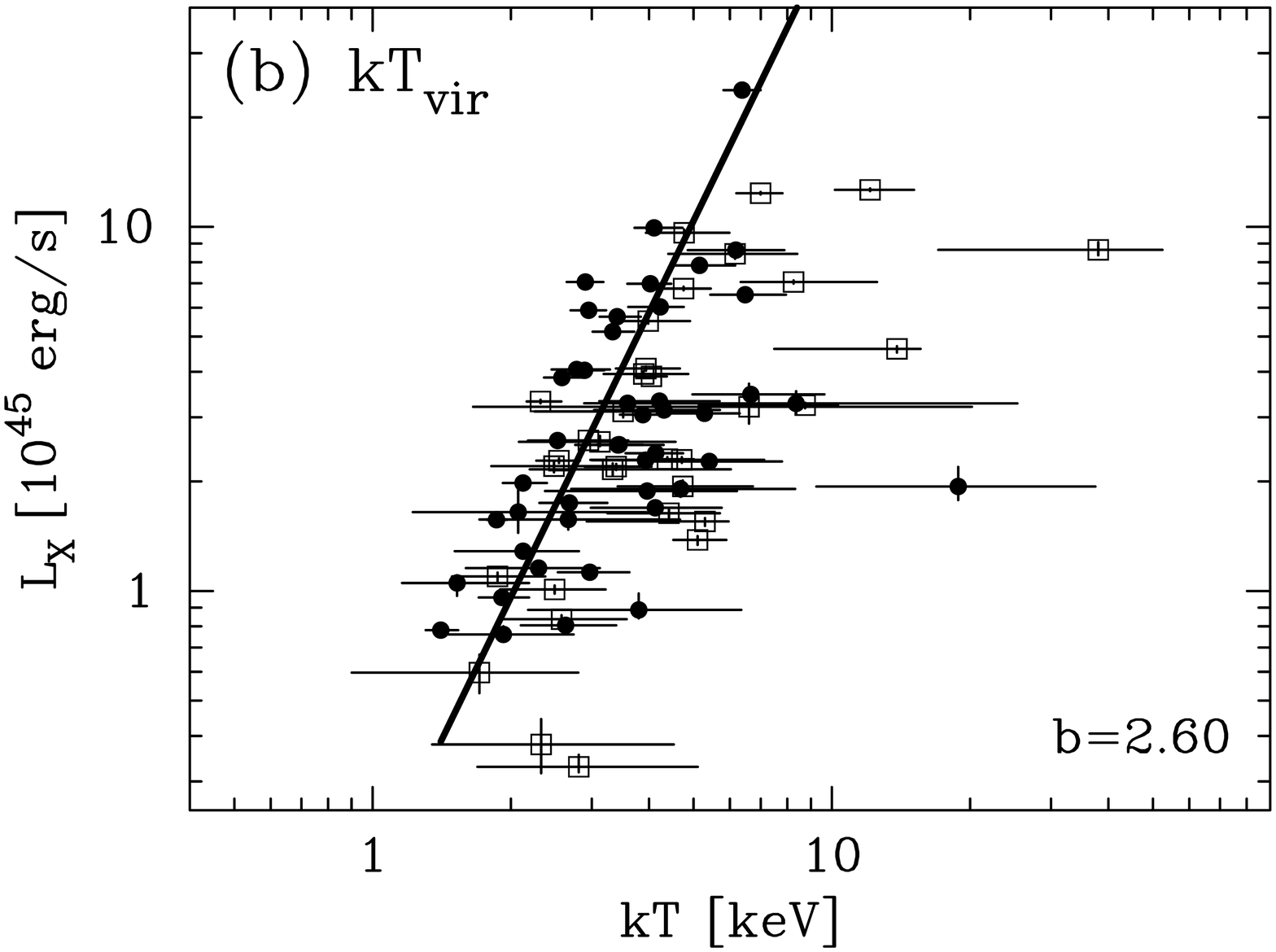}
\end{center}
\caption{$L_X-T$ relation for (a) $T_X$ and (b) $T_{\rm vir}=\beta T_Xx_{\rm vir}^2/(1+x_{\rm vir}^2)$.  The filled circles and open squares represent the regular and irregular clusters, respectively (see section \ref{ss:Xsym}). Each solid line represents the best-fit relation in the form $L_X = a T^b$ with $a$ and $b$ being parameters.
}
\label{fig:Lx-T}
\end{figure*}

Equation~(\ref{betaTiso-Tvir}) or (\ref{eq:Tvir-Tiso}) suggests that the $L_X - T_X$ relation, which is obtained from X-ray observations, would differ from $L_X - T_{\rm vir}$.  The $L_X-T$ relations are compared in figure~\ref{fig:Lx-T}, where $T=T_X$ in figure~\ref{fig:Lx-T}a and $T=T_{\rm vir}=\beta T_Xx_{\rm vir}^2/(1+x_{\rm vir}^2)$ in figure~\ref{fig:Lx-T}b. 

For $T=T_X$, we obtain the best-fit relation and $90\%$ confidence levels $L_X=(1.48_{-0.36}^{+0.52})\times 10^{-2}(kT_X)^{3.02_{-0.17}^{+0.15}}$ with correlation coefficient $r_{xy}=0.49$. On the other hand, for $T=T_{\rm vir}$ given by equation~(\ref{eq:Tvir-Tiso}) we obtain less steep slope $L_X=(1.59_{-0.41}^{+0.46})\times 10^{-1}(kT_{\rm vir})^{2.60_{-0.24}^{+0.21}}$ with $r_{xy}=0.37$. It should be noted that many of clusters deviating significantly from the best-fit $L_X-T_{\rm vir}$ relation are those of which surface brightness profiles are far from spherical symmetry. The redshifts of the clusters are not taken into account for the analysis, because it is suggested from observations that the redshift dependence of $L_X-T$ is weak (e.g., \cite{Fairley00}).

\section{Observational Implications}\label{sec:observation}

We examine 121 clusters: 42 nearby clusters of $z=$~0.011--0.097 (\cite{Mohr99}) and 79 distant clusters of $z=$~0.10--0.82 (\cite{OtaD}; \cite{OM02}; \cite{OM04}) with cosmological parameters ${\rm \Omega _m}=0.3$, ${\rm \Omega _\Lambda}=0.7$ and $H_0=70~{\rm km~s^{-1}~Mpc^{-1}}$. \citet{OM02} mentioned the selection effect due to the detection limit of {\it ASCA}. Most of the clusters concerned here have redshifts $z<0.5$ where the selection effect is weak; the numbers of clusters of $z>0.5$ are 2 out of 50 and 6 out of 71 for the smaller and larger core group (see below), respectively.

\begin{figure}
\begin{center}
\FigureFile(85mm,60mm){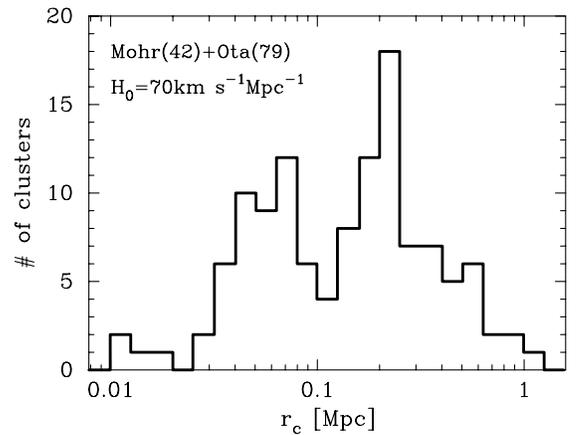}
\end{center}
\caption{ Distribution of core radii of 121 clusters.  One can see two groups of the core radii, which have peaks in 40--80~kpc and 140--300~kpc, respectively. }
\label{fig:4}
\end{figure}

The 121 clusters can be classified into two groups: ``smaller core" that is smaller than 120~kpc and ``larger core" larger than that, where 120~kpc corresponds to the ``valley" in the distribution of core radii (figure~\ref{fig:4}). Some clusters fit better to the double $\beta$-model, i.e. superposition of two single $\beta$-models. For such clusters, we apply the double $\beta$-model and choose the core that is dominant in the emissivity-averaged temperature. In other words, we take the component that dominates the luminosity of the cluster.

In the following subsections, we discuss the possible origin of the typical two core sizes, considering the $r_c-r_{\rm vir}$ relation, presence of cD galaxy, asymmetry in X-ray surface brightness and effect of radiative cooling.

\subsection{Relation between Virial Radius and Core Radius}\label{ss:virial}

\begin{figure}
\begin{center}
\FigureFile(85mm,60mm){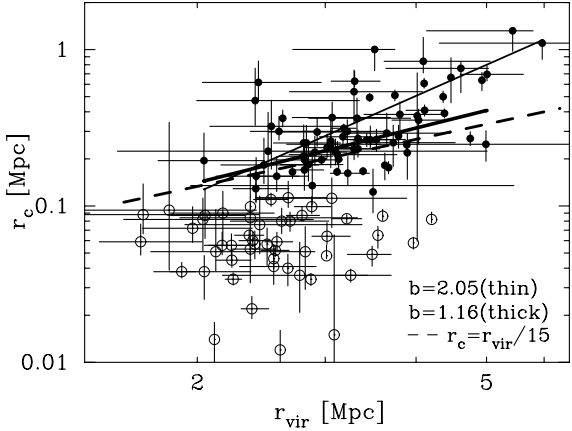}
\end{center}
\caption{$r_c-r_{\rm vir}$ relation. The open and filled circles represent the clusters of smaller and larger core group, respectively. The thin and thick solid lines represent the best-fit relation ($r_c=a r^b_{\rm vir}$) for the larger core group without statistical errors and that for the clusters of $r_c =$~140--300~kpc of the \citet{OM02} sample, respectively. The dashed line represents $r_{\rm c}=r_{\rm vir}/15$ (see section~\ref{sec:simulation}).
}
\label{fig:5}
\end{figure}

We assume that clusters formed by spherical collapse and virialized at $z_{\rm vir}$ are observed at $z_{\rm obs} = z_{\rm vir}$. We define $r_{\rm vir}$ as a radius at which the average density is equal to $\Delta _c$ times the critical density at $z_{\rm vir}$, i.e. $\bar{\rho}(r_{\rm vir})=\Delta _c\rho _{\rm crit}(z_{\rm vir})$ with $\Delta _c=18\pi^2{\rm \Omega_m}(z_{\rm vir})^{0.427}$ (\cite{NS97}). For $r_{\rm vir}$ thus determined, we examine its correlation with $r_c$ for the 121 clusters (figure~\ref{fig:5}).

For the larger core group, we obtain the best-fit relation $r_c=2.95\times 10^{-2} r_{\rm vir}^{2.05}$ with correlation coefficient $r_{xy}=0.54$; here, statistical errors are ignored, since the errors of $r_{\rm vir}$ are not available for the \citet{Mohr99} sample.
If restricted to the \citet{OM02} sample for which the errors are available, we obtain $r_c=(7.94_{-4.71}^{+7.19})\times 10^{-3} r_{\rm vir}^{3.06_{-0.54}^{+0.76}}$ with $\chi^2/{\rm d.o.f.}=59.03/43$.
 
However, it should be noted that the $r_c$ distribution of the larger core group exhibits a tail extended to $\sim 1$~Mpc (figure~\ref{fig:4}), and the $r_c - r_{\rm vir}$ relation likely depends much on the range of $r_c$ to be taken into account for the analysis.  Actually, for the clusters in the narrower $r_c =$~160--250~kpc and broader $r_c =$~120--400~kpc ranges, we obtain $r_c=(1.38_{-0.91}^{+1.25})\times 10^{-1} r_{\rm vir}^{0.39_{-0.56}^{+0.97}}$ with $\chi^2/{\rm d.o.f.}=4.03/13$ and $r_c=(1.91_{-1.39}^{+2.56})\times 10^{-2}~r_{\rm vir}^{2.27_{-0.74}^{+1.16}}$ with $\chi^2/{\rm d.o.f.}=26.90/32$, respectively.  For the intermediate range $r_c =$~140--300~kpc, we obtain $r_c=(6.31_{-3.74}^{+18.24})\times 10^{-2}~r_{\rm vir}^{1.16_{-0.79}^{+1.21}}$ with $\chi^2/{\rm d.o.f.}=17.25/24$.  From this result, we may consider that the clusters of the larger core group marginally satisfy the self-similar relation \citep{OM04}. Assuming the self-similarity $r_c \propto r_{\rm vir}$ for the clusters of $r_c =$~140--300~kpc, we obtain the best-fit relation $r_c=r_{\rm vir}/16.6$.

On the other hand, for the smaller core group, we find no clear correlation between $r_c$ and $r_{\rm vir}$.  This may imply that the origin of small cores is not explained by simple self-similar collapse.

\subsection{Relation between cD Galaxy and Core Radius}\label{ss:cd-obs}

cD galaxy is a giant elliptical galaxy which is located near the gravitational center of a cluster, and its mass reaches $10^{11-12}~\MO$ which is comparable to the sum of other member galaxies. In our sample, 50 out of 121 clusters have central cDs or giant ellipticals.  We classify the clusters by BM-type (\cite{BM70}) into cD cluster (Type I and I-II) and others (Type II, II-III and III), using NASA/IPAC EXTRAGALACTIC DATABASE, and find 75 BM-types: 38 in nearby and 37 in distant clusters.  Although there is some question about the accuracy of the BM-type classification, as far as a cluster possesses one and two brightest galaxies, the cluster can safely be assigned to Type I and I-II, respectively (see e.g., \cite{LB77}).

We classify the 75 clusters further into the smaller and larger core groups by $r_c \sim 120$~kpc described above.  For each group, the fraction of clusters classified into cD clusters is shown in table~\ref{ta:1}a.  The fraction of cD cluster is large in nearby clusters compared to the distant ones in both the smaller and larger core groups.  This could be due to an observable limit of the surface brightness for the faint envelopes of distant cD galaxies.

As for the nearby clusters, 73\% of the smaller core group and 52\% of the larger core group are classified into cD clusters; no distinctive difference is found between the two groups.  The 12 (52\%) Type I and I-II clusters of the larger core group evidently have dominant cD galaxies or giant ellipticals; cD galaxies are not characteristic of the smaller core clusters.  Therefore, we may say that the origin of small cores are not always ascribed to the presence of cD galaxies.

\begin{table*}
\caption{The fractions of cD clusters and ``regular" clusters.}
\label{ta:1}
\begin{center}
\begin{tabular}{lcrrrrrr}
\hline\hline
\noalign{\smallskip}
& Classification & \multicolumn{2}{c}{Nearby} & \multicolumn{2}{c}{Distant} & \multicolumn{2}{c}{Total} \\
\noalign{\smallskip}
& Clusters in the smaller core group & 16/42 & (38\%) & 34/79 & (43\%) & 50/121 & (41\%)\\
\noalign{\smallskip}
\hline
\noalign{\smallskip}
\noalign{\smallskip}
(a) & cD clusters in the smaller core group of BM ..& 11/15 & (73\%) & 2/9 & (22\%) & 13/24 & (54\%)\\
\noalign{\smallskip}
& cD clusters in the larger core group of BM ....&12/23 & (52\%) & 4/28 & (14\%) & 16/51 & (31\%)\\
\noalign{\smallskip}
\noalign{\smallskip}
(b) & Regular clusters in the smaller core group .....& 10/15 & (67\%) & 28/34 & (82\%) & 38/49 & (78\%)\\
\noalign{\smallskip}
& Regular clusters in the larger core group .......& 15/24 & (63\%) & 17/45 & (38\%) & 32/69 & (46\%)\\
\noalign{\smallskip}
\hline
\noalign{\smallskip}
\multicolumn{8}{l}{(a) 75 clusters are classified into the smaller and larger core groups.}\\
\multicolumn{8}{l}{(b) 118 clusters are classified into the smaller and larger core groups.}
\end{tabular}
\end{center}
\end{table*}

\subsection{Relation between X-ray Symmetry and Core Radius}\label{ss:Xsym}

The $\beta$-model-based core size may be affected by asymmetry in the X-ray surface brightness.  Such asymmetry can be due to the presence of radio lobes or cold clumps in the core, or caused by mergers.  

We classify the 121 clusters into ``regular" and ``irregular" groups simply by the surface brightness symmetry.  The classification of distant clusters is referred to \citet{OtaD}, who examines the deviation of the surface brightness centroid varying the aperture radius; if the deviation is greater than the 3$\sigma$ level, the cluster is classified into irregulars. For 19 nearby clusters, following \citet{Mohr95} (see also \cite{Mohr93}), we classify clusters with $w_{\vec{x}}/2 > 3\sigma_w$ into irregulars. For other 20 nearby clusters, following \citet{BT96}, we classify clusters with $P_1^{\rm (pk)}/P_0^{\rm (pk)} > 3\sigma$ into irregulars, where $\sigma$ is estimated from the 90\% confidence in their Table 2. Since the data are not available for three nearby clusters, we investigate 39 nearby and 79 distant clusters, i.e. 118 clusters in total.

The fractions of regular clusters are shown for the smaller and larger core groups in table~\ref{ta:1}b.  The regular fraction is larger in the smaller core group of the distant clusters, while the fraction is almost the same between the smaller and larger core groups for the nearby clusters. The difference between the distant and nearby clusters may be due to mergers or be interpreted as that distant clusters are well relaxed compared to nearby ones, though there could be an observational effect such that detection of substructures in distant clusters are difficult.  As for the core sizes, however, no clear difference is found in the nearby clusters.  Therefore, it is unlikely that X-ray asymmetry is characteristic to small core clusters.

\subsection{Relation between Radiative Cooling and Core Radius}\label{ss:cool}

In the cores of some clusters, the gas undergoes significant radiative cooling, and then inflow of the ambient gas may occur to compensate the pressure loss inside (e.g., \cite{Fabian94}).  This process forms a peaked density profile responsible for the smaller cores.  

\begin{figure*}
\begin{center}
\FigureFile(80mm,60mm){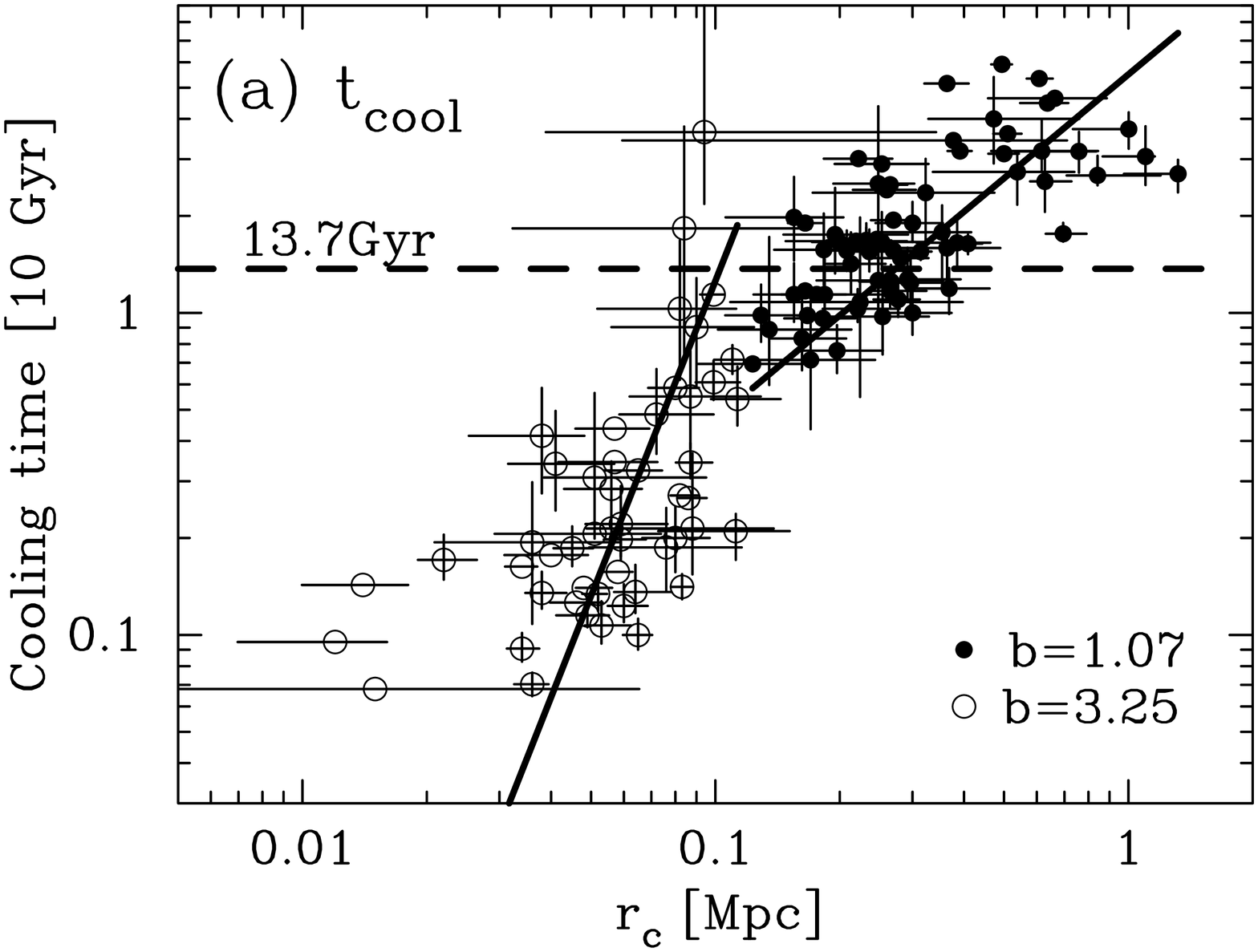}
\FigureFile(80mm,60mm){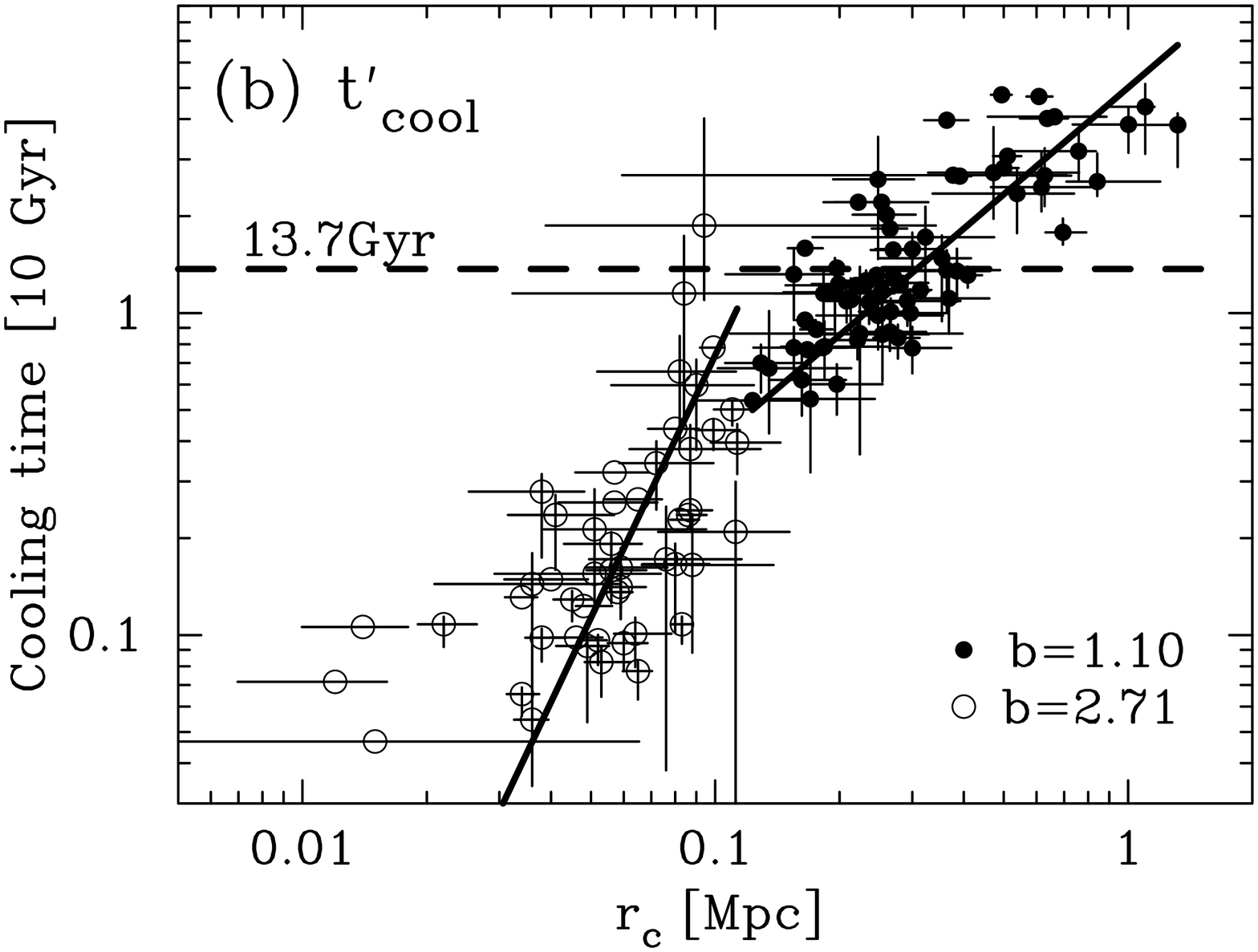}
\end{center}
\caption{ Relation of $r_c$ with (a) $t_{\rm cool}$ and with (b) $t_{\rm cool}'$. The solid lines represent the best-fit relation, $t_{\rm cool}$ (or $t_{\rm cool}'$) $= a r_c^b$, for the smaller core group (open circles) and for the larger core group (filled circles). Statistical errors are not taken into account for the analysis, since the errors of cooling time are not available for the \citet{Mohr99} sample. The dashed line represents the Hubble time $t_{\rm H}$.
}
\label{fig:6}
\end{figure*}

The cooling time at the cluster center due to bremsstrahlung is written as
\begin{equation}\label{eq:tcool}
t_{\rm cool}=\frac{3k\sqrt{T_X}}{q_{\rm ff}n_{e0}},
\end{equation}
where $n_{e0}$ is the electron number density and $q_{\rm ff}$ is a coefficient for the emissivity $q_{\rm ff}\sqrt{T}$.  In figure~\ref {fig:6}a is shown $t_{\rm cool}$ against the core radius $r_c$ for the 121 clusters.  The best-fit relation is obtained as $t_{\rm cool}=2.24\times 10^3~r_c^{3.25}$ with $r_{xy}=0.44$ for the smaller core group and $t_{\rm cool}=5.50~r_c^{1.07}$ with $r_{xy}=0.55$ for the larger core group.

In Section 2, we argue that $T_X$ is not always a good quantity to represent $T_{\rm vir}$ in the context of the $\beta$-model.  So that, we define another cooling time 
\begin{equation}\label{eq:tcool'}
t_{\rm cool}'=\frac{3k\sqrt{T_{\rm vir}}}{q_{\rm ff}n_{e0}}
=\left(\beta\frac{x_{\rm vir}^2}{1+x_{\rm vir}^2}\right)^{1/2}t_{\rm cool},
\end{equation}
which depends on $\beta$.  The cooling time $t_{\rm cool}'$ is shown in figure~\ref {fig:6}b.   We obtain the best-fit relation $t_{\rm cool}'=3.80\times 10^2~r_c^{2.71}$ with $r_{xy}=0.51$ for the smaller core and $t_{\rm cool}'=5.01~r_c^{1.10}$ with $r_{xy}=0.75$ for the larger core groups.  The relation of $t_{\rm cool}'$ with $r_c$ is clearer than that of $t_{\rm cool}$ for the larger core clusters which have long cooling time.

In comparison with the larger core group, the relation of $r_c$ with cooling time is not clear in the smaller core group. As discussed in section~\ref{ss:virial}, neither the self-similar relation between $r_c$ and $r_{\rm vir}$ is clear in the smaller core group. These may suggest that the clusters of small cores are affected by radiative cooling after the self-similar collapse.

Also cDs or galaxies in the central region can alter the core structure through their AGN activities, supernovae feedback or galactic winds.  However, as discussed in section 3.2, the difference in the relation with such galaxies is not clear between the smaller and larger core groups as compared to the relation with the cooling time.  Moreover, heat or pressure due to such activities of central galaxies likely works against the core getting smaller, while the relevant metal enrichment enhances radiative cooling.

If the gas-mass fraction in the cooling core is nearly constant among the clusters, a scaling law follows the self-similarity $r_{\rm vir}\propto r_c$, as (see section~\ref{sec:theory} and equation~(\ref{eq:ng0}))
\begin{equation}\label{eq:tcool'scale1}
t_{\rm cool}'\propto T_{\rm vir}^{1/2}n_{e0}^{-1}
\propto r_c^2r_{\rm vir}^{-1} \propto r_c.
\end{equation}
This dependence is seen in figure~\ref {fig:6}b for the clusters of the larger core group.  For large core clusters of $t'_{\rm cool}\gtrsim t_{\rm H}$, radiative cooling is not yet significant and their cores remain to follow the self-similar relation.   Otherwise, if cooling is significant, the core would evolve to deviate from the self-similarity.  This is likely the case for the smaller core group: no clear correlations between $r_c$ and $r_{\rm vir}$ in figure~\ref {fig:5}.

\begin{figure*}
\begin{center}
\FigureFile(80mm,60mm){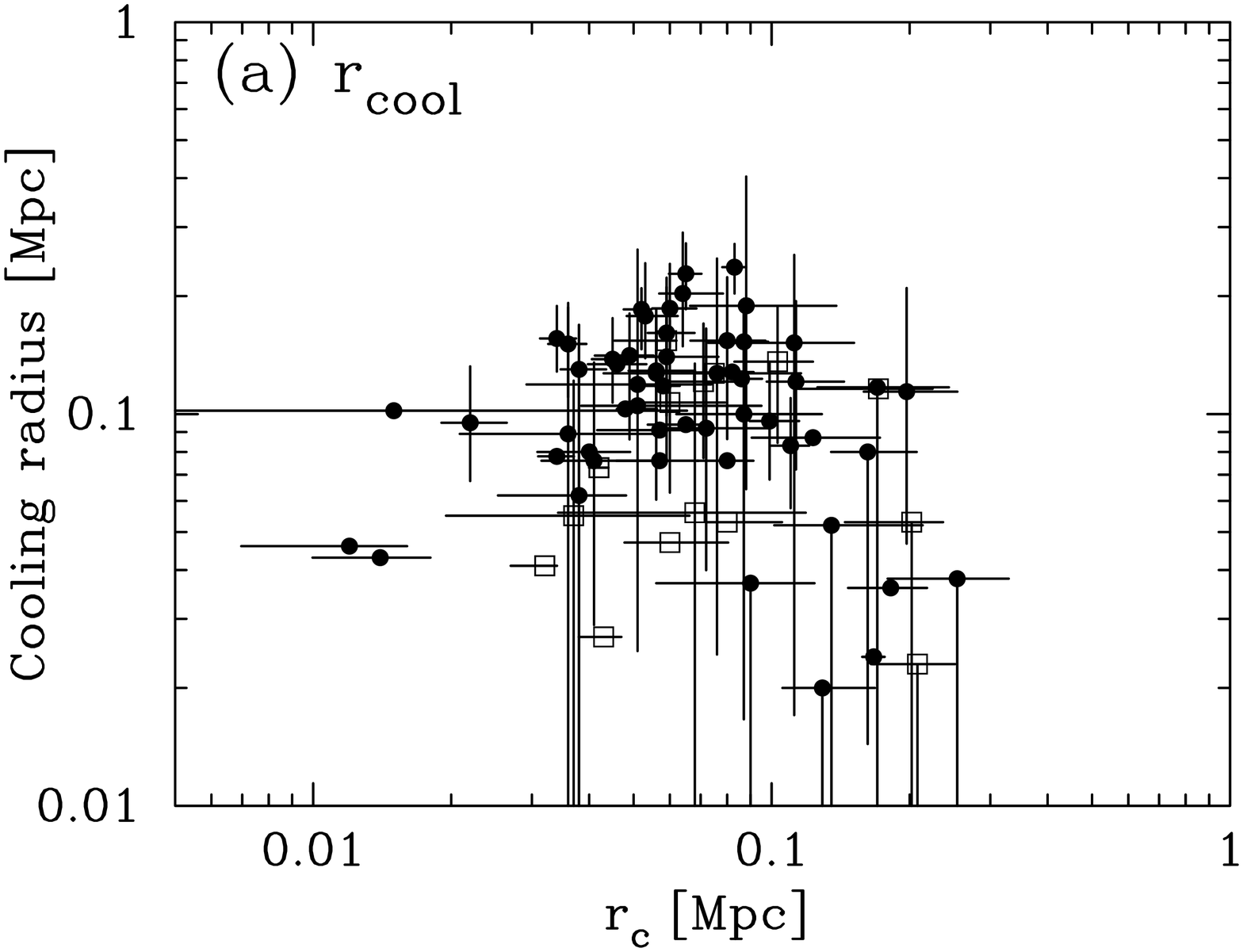}
\FigureFile(80mm,60mm){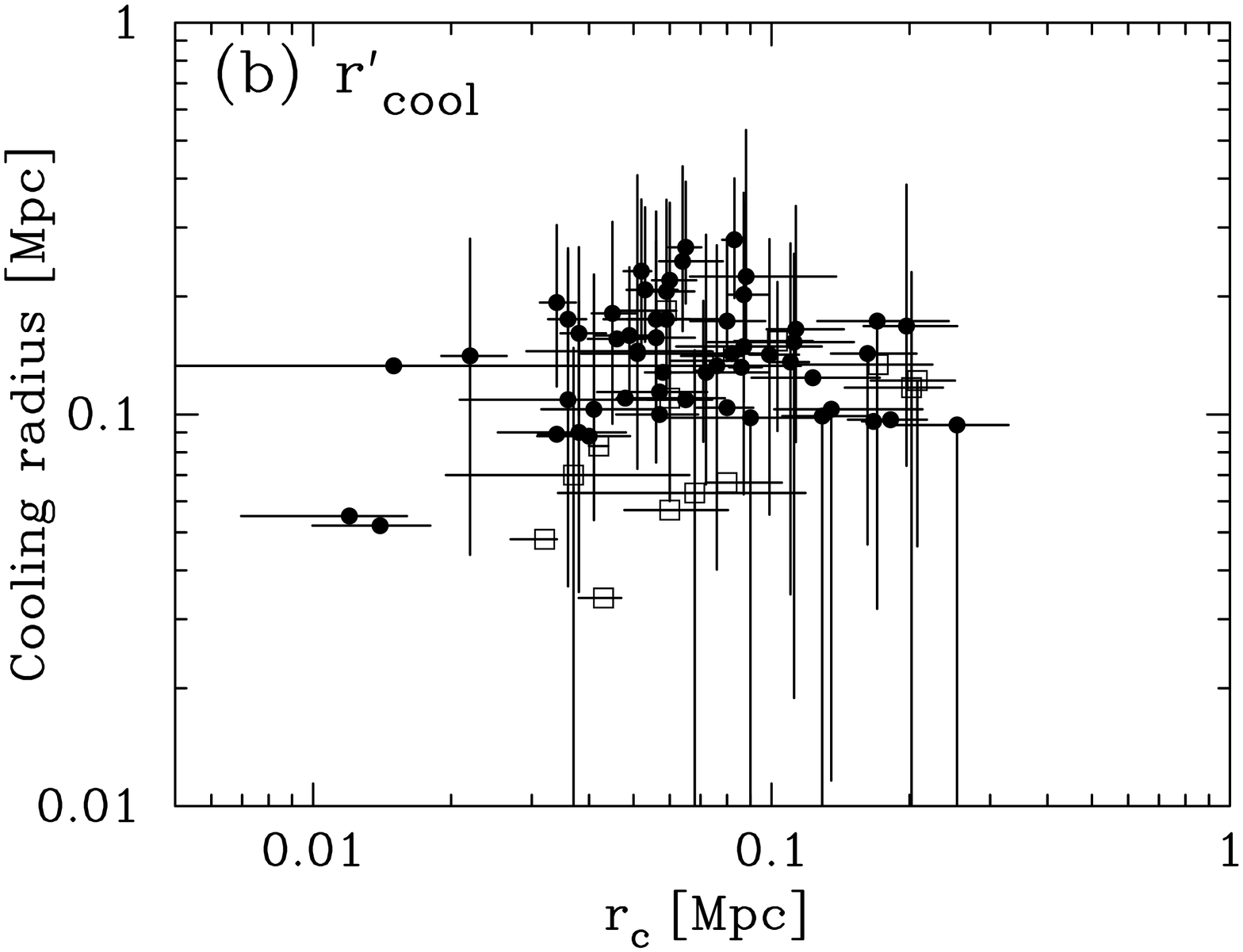}
\end{center}
\caption{ Relation of $r_c$ with (a) $r_{\rm cool}$ and with (b) $r_{\rm cool}'$. For a double $\beta$-model cluster, each core radius is plotted: the filled circle and open square represent the dominant core and another (minor) one, respectively.  The single $\beta$-model clusters are represented by the filled circles.
}
\label{fig:7}
\end{figure*}

In order to see the effect of cooling on the smaller core group clusters, we define the cooling radius based on the $\beta$-model, as
\begin{eqnarray}\label{eq:rcool}
r_{\rm cool} &=& r_c \left({t_{\rm cool,10}}^{-2/3\beta} -1 \right)^{1/2} \\ \mbox{or}\nonumber\\
r_{\rm cool}' &=& r_c \left({t_{\rm cool,10}'}^{-2/3\beta} -1 \right)^{1/2}
\end{eqnarray}
for ${t_{\rm cool,10}}$ or ${t'_{\rm cool,10}}$ in units of 10~Gyr, and investigate 70 clusters of $t_{\rm cool,10} < 1$.  The cooling radius is plotted against the core radius in figure~\ref {fig:7}.  No clear correlations are seen in $r_{\rm cool}-r_c$ or $r'_{\rm cool}-r_c$.  

If $t'_{\rm cool}$ is significantly shorter than $t_{\rm H}$, we have an approximate relation
\begin{equation}\label{eq:rcool'scale}
r_{\rm cool}' \propto r_c {t_{\rm cool}'}^{-1/3\beta} \propto r_c^{1-2.7/3\beta},
\end{equation}
where, in the last expression, $t_{\rm cool}'\propto r_c^{2.7}$ suggested from figure~\ref{fig:6}b is taken into account, though the correlation is quite weak.  Consequently, for various values of $\beta$ from $\sim 0.6$ to $\sim 1$ (figure~\ref{fig:1}), $r_c$ dependence of $r'_{\rm cool}$ would be smeared and the self-similar relation might be lost for the smaller core group.  This suggests that cooling in the core is likely connected with the origin of small cores of clusters which are out of self-similarity. 

\section{Hydrodynamical Approach}\label{sec:simulation}

Using an SPH (Smoothed Particle Hydrodynamics) code based on \citet{HK89} and \citet{Monaghan92}, we simulate the $\beta$-model and investigate the behavior of the isothermal intracluster gas in order to make sure the arguments in the previous sections.  Our main concern here is the self-similar relation $r_c\propto r_{\rm vir}$ and the effect of the potential due to cD galaxy on the isothermal gas without cooling; the effect of radiative cooling will be discussed in a separate paper.  Table~\ref{ta:2} shows the resolution and systematic error estimated for the calculations.

\begin{table}
\caption{Resolutions $h$ and systematic errors $O(h^2)$.}
\label{ta:2}
\begin{center}
\begin{tabular}{p{0.2\linewidth}cccc}
\hline\hline
\noalign{\smallskip}
$N_{\rm SPH}$ & $h_{0.01^*}$ & $h_{0.05^*}$ & $h_{0.1^*}$ & $O(h^2)$  \\
\noalign{\smallskip}
& (kpc) & (kpc) & (kpc) & (\%) \\
\noalign{\smallskip}
\hline
\noalign{\smallskip}
30000 & 31.5 & 18.4 & 14.6 & 0.961 \\
50000 &26.6 & 15.6 & 12.3 & 0.705 \\
\noalign{\smallskip}
\hline
\noalign{\smallskip}
\multicolumn{5}{l}{$^*$ The gas density $n_g$ (${\rm cm^{-3}}$).}
\end{tabular}
\end{center}
\end{table}

We consider that a cluster consists of galaxies, gas and dark matter with the mass ratio $M_{\rm gal}:M_{\rm gas}:M_{\rm DM}=1:5:30$.  The Momentum equation for the gas is written as
\begin{equation}\label{eq:momentum}
\frac{dv_i}{dt}=-\frac{1}{\rho _i}
\nabla P_i+a_i^{\rm visc}-\nabla\Phi _i-G_i,
\end{equation}
where the first and third terms on the right hand side represent the force due to the pressure gradient and self-gravity, respectively.  The second term is the artificial viscosity, for which we adopt the Monaghan-Gingold value (\cite{MG83}; see also \cite{HK89}). The last term represents the force due to the components other than the gas, i.e. the dark matter and member galaxies,
\begin{equation}\label{eq:external}
-G_i=-\frac{GM_{\rm ex}(r_i)}{r_i^2}
\frac{r_i}{\mid r_i\mid },
\end{equation}
where $M_{\rm ex}(r_i)$ is the mass of the components within radius $r_i$. For their distributions, we take an approximate King profile
\begin{equation}\label{eq:King}
\rho_{\rm K}(r)=\rho_{{\rm K}0}[1+(r/r_{\rm K})^2]^{-3/2}
\end{equation}
with $\rho _{\rm K}(r>r_{\rm vir})=0$, where $\rho_{{\rm K}0}$ is given for each component according to the mass ratio and $r_{\rm K}$ is taken to be the initial core size of gas.

For the initial distribution of the gas we set the $\beta$-model with $\beta=1$, $n_{g0}=0.04$~${\rm cm^{-3}}$ and $r_{\rm vir}/r_c=x_{\rm vir}=15$ (see figure~\ref{fig:5}), and calculate the evolution to be hydrostatic with the gas being isothermal.  The initial value of the gas temperature $T$ is taken to be $T_{\rm vir}$, considering $\beta x_{\rm vir}^2/(1+x_{\rm vir}^2) = 0.996 \simeq 1$ at the initial state.

\subsection{Comparison with the $\beta$-model}\label{ss:betafit}

We carry out calculations for four clusters of the virial radius 2.3, 3.0, 3.6 and 4.0~Mpc until 3.0~Gyr, which is enough longer than the free fall time $\sim 1.2$~Gyr for $n_{g0}=0.04$~${\rm cm^{-3}}$, and examine the resultant intracluster gas.  We find that the gas profile is well reproduced by the $\beta$-model with the parameter values shown in table~\ref{ta:3}.  The resultant $\beta$ is smaller than unity (initial value), yet the relation $T_{\rm vir} \simeq \beta T$ is attained for the gas temperature $T$, as expected from the argument in section~\ref{sec:theory}.

\begin{figure*}
\begin{center}
\FigureFile(80mm,60mm){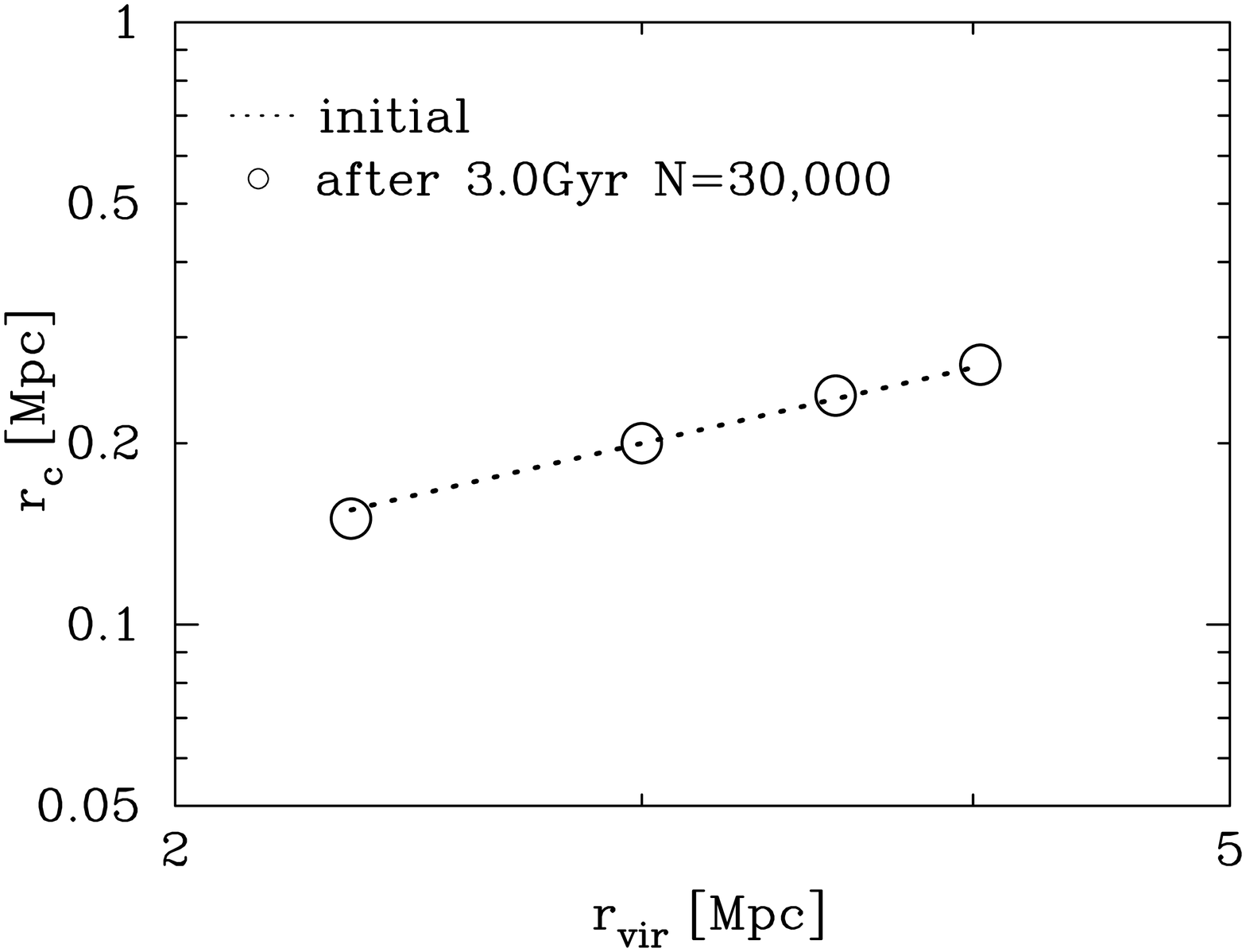}
\FigureFile(80mm,60mm){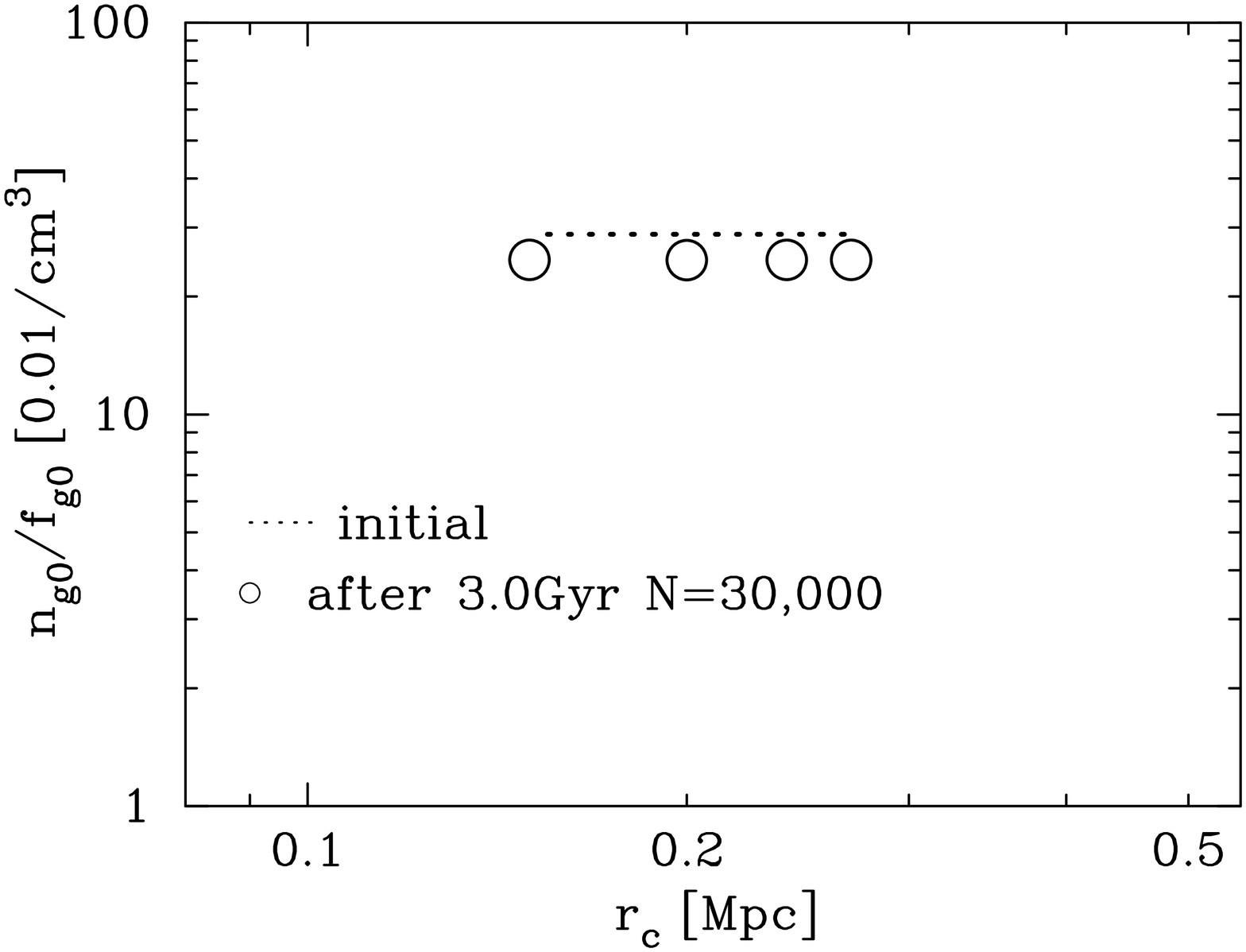}
\end{center}
\caption{
(a) $r_c-r_{\rm vir}$ and (b) $n_{g0}/f_{g0}-r_c$ relation. The dotted lines represent the initial values: (a) $r_c=r_{\rm vir}/15$ and (b) $n_{g0}/f_{g0}=0.04/0.139$~${\rm cm^{-3}}$.  The circles represent the results after 3.0~Gyr with $N = 30000$ SPH-particles.}
\label{fig:relation-sph}
\end{figure*}

Also the self-similar relation between $r_c$ and $r_{\rm vir}$ is kept in spite of that $\beta$ becomes smaller than unity, as shown in figure~\ref{fig:relation-sph}a.  This implies that the core radius well reflects the core of the gravitational potential consistently with the $\beta$-model.  The consistency can be examined in another way as follows.  
When the gas follows the $\beta$-model, the gas density at the cluster center is written as
\begin{equation}\label{eq:ng0}
n_{g0}=\frac{9k \beta T f_{g0}}{4\pi Gr_c^2},
\end{equation}
where $f_{g0}$ is the gas-mass fraction at the cluster center. If the self-similar relation $r_c \propto r_{\rm vir}$ is attained, $T_{\rm vir}/r_c^2 \simeq \beta T/r_c^2$ becomes independent of $r_c$ from equation (\ref{eq:virialtheorem}) and $M_{\rm vir} \propto r_{\rm vir}^3 \propto r_c^3$, which is obtained also from the calculations (see table~\ref{ta:3}).  Accordingly, $n_{g0}/f_{g0}$ should be independent of $r_c$, if the calculated structure is consistent with the $\beta$-model.  This behavior is seen in figure~\ref{fig:relation-sph}b, where $n_{g0}/f_{g0}$ is plotted against $r_c$ for the four calculations. 

\begin{table}
\caption{Best-fit parameters of the $\beta$-model.}
\label{ta:3}
\begin{center}
\begin{tabular}{cccccccc}
\hline\hline
\noalign{\smallskip}
\multicolumn{3}{c}{Model}&&\multicolumn{3}{c}{After 3.0Gyr}\\
\cline{5-7}
\noalign{\smallskip}
$M_{\rm vir}$ & $r_{\rm vir}$ & $r_c$ && $r_c$ & $n_{g0}$ & $\beta$ \\
($10^{15}\MO$) & (Mpc) & (kpc) && (kpc) & $({\rm cm^{-3}})$ &\\
\noalign{\smallskip}
\hline
\noalign{\smallskip}
0.48 & 2.30 & 150 && 150 & 2.00 & 0.83\\
1.03 & 3.00 & 200 && 200 & 2.04 & 0.86\\
1.72 & 3.60 & 240 && 240 & 2.03 & 0.87\\
2.50 & 4.00 & 270 && 270 & 2.10 & 0.88\\
\noalign{\smallskip}
\hline
\noalign{\smallskip}
\end{tabular}
\end{center}
\end{table}

\subsection{cD galaxy}\label{ss:cd-hydro}

\begin{figure}
\begin{center}
\FigureFile(85mm,60mm){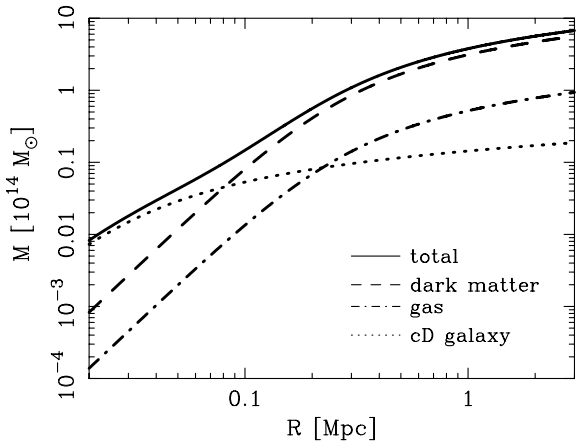}
\end{center}
\caption{
Initial profile of the integrated mass: dark matter (dashed line), gas (dot-dashed line), cD galaxy (dotted line), and total (solid line). 
}
\label{fig:8}
\end{figure}

Intending to estimate the maximal effect of a central cD galaxy, we take a large cD galaxy with the core radius $r_{\rm K}=20$~kpc into account for the cluster potential. We consider a cluster of $r_{\rm vir}=3$~Mpc, $r_c=200$~kpc and $n_{g0}=0.04$~${\rm cm^{-3}}$ and simply replace the galaxy component by the cD one so as for $r_{\rm vir}$, $M_{\rm vir}$, $T_{\rm vir}$ and the gas-mass fraction to be the same values as those in the case without the cD for the same mass ratio; this also allows to keep the resolution for the computation resources.  Although the potential due to other member galaxies is ignored here, we make sure that their contribution is negligibly small compared to other components. The integrated mass of each component is shown in figure~\ref{fig:8}.

\begin{figure}
\begin{center}
\FigureFile(85mm,60mm){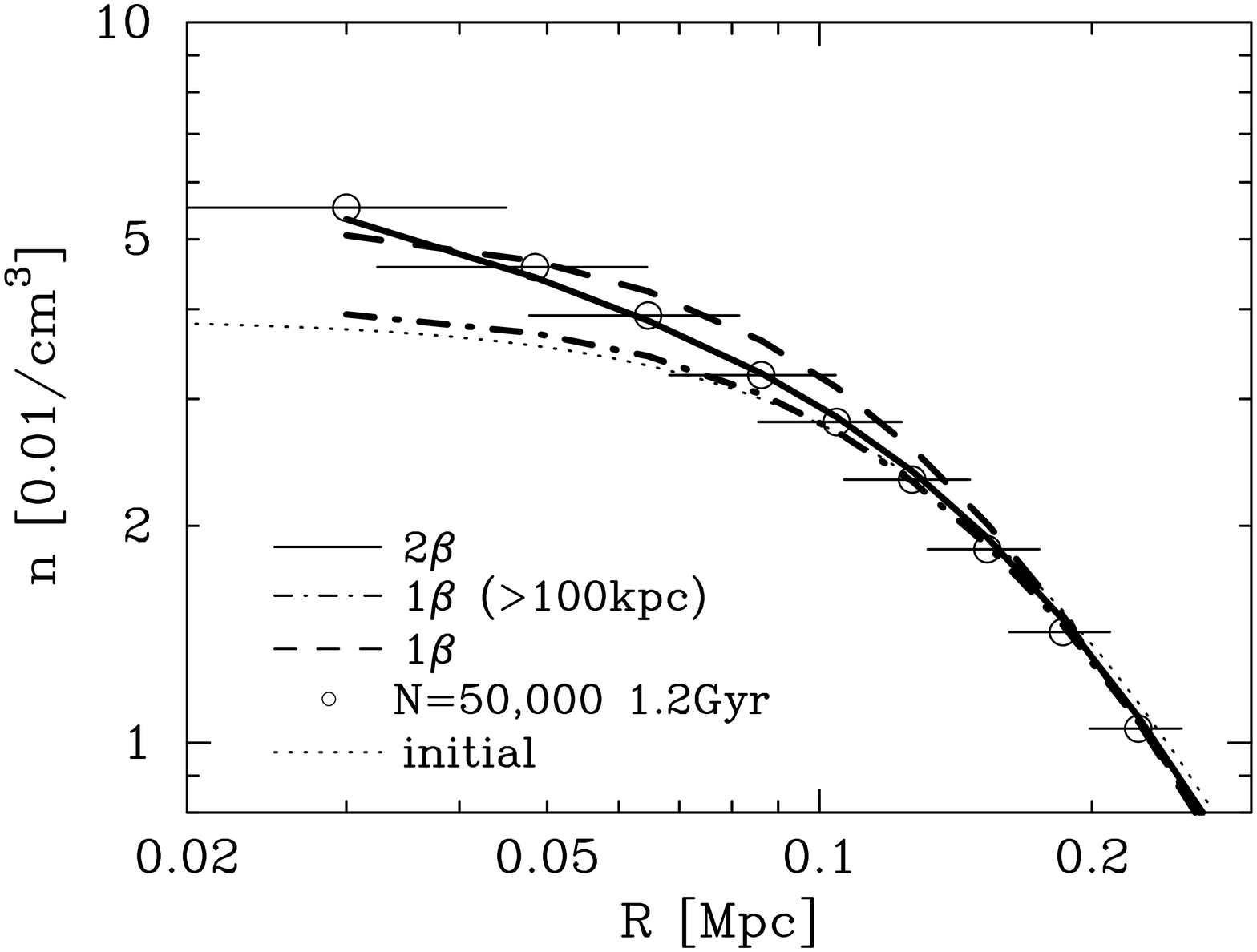}
\end{center}
\caption{Gas density profile: the circles represent the profile after 1.2~Gyr with $N=50000$ SPH-particles and the bars attached represent the errors (smoothing lengths).  The lines represent the best-fit single and double $\beta$-models.
}
\label{fig:9}
\end{figure}

Figure~\ref{fig:9} shows the gas profile after 1.2~Gyr with $N = 50000$ SPH-particles.  The density increases toward the cluster center in comparison with the initial profile. The best-fit $\beta$-model (see table~\ref{ta:4}) to the resultant profile gives $r_c=150$~kpc smaller than the initial core size, while the profile at $> 100$~kpc varies little from the initial one.  This means that the presence of a central cD galaxy appears merely as an excess within 100~kpc and little affects the outer region.  Therefore, we try to fit the gas profile using the double $\beta$-model as follows.

We first obtain the best-fit parameters $n_{g0}$, $r_c$ and $\beta _1$ for the outer component, $r=$~0.1--2.0~Mpc, and then include the inner component to obtain $n_{g0,2}$ and $r_{c,2}$ with $n_{g0}$, $r_c$ and $\beta _1$ fixed, where we assume $\beta _2=\beta _1$ as done in the double $\beta$-model analyses by \citet{Mohr99} and \citet{OM02}.  The inner core radius thus obtained, $r_{c,2}=41$~kpc, is about twice the core radius $r_{\rm K} = 20$~kpc of the cD galaxy.  The extended core of the gas compared to the cD potential is likely due to the pressure gradient.  As seen in figure~\ref{fig:4}, the core radius $\sim 40$~kpc is comparable to the smaller end of the smaller core group, but is too small to explain the distribution 40--80~kpc because the typical core size of cD galaxies is less than $\sim 20$~kpc (see e.g., \cite{SG00}).

\begin{table*}
\caption{Best-fit parameters of the single and double $\beta$-models.}
\label{ta:4}
\begin{center}
\begin{tabular}{ccccccccc}
\hline\hline
\noalign{\smallskip}
&&\multicolumn{3}{c}{Single $\beta$-model}&&\multicolumn{3}{c}{Double $\beta$-model$^a$}\\
\cline{3-5}\cline{7-9}
\noalign{\smallskip}
\noalign{\smallskip}
&& $r_c$ & $n_{g0}$ & $\beta _1$ && $r_{c,2}$ & $n_{g0,2}$ & $\beta _2$\\
Model && (kpc) & $({\rm cm}^{-3})$ & && (kpc) & $({\rm cm}^{-3})$ & \\
\noalign{\smallskip}
\hline
\noalign{\smallskip}
1$\beta$ ...................... && 150 &  5.27 & 0.91&&&& \\
1$\beta$ ($>$100~kpc) ...&& 181 &  4.04 & 0.95 &&&& \\
2$\beta$ ...................... &&...&...&...&& 41 & 2.66 &...\\
\noalign{\smallskip}
\hline
\noalign{\smallskip}
\multicolumn{9}{l}{$^a$ Parameters of the outer component are fixed at the values obtained}\\
\multicolumn{9}{l}{outside 100~kpc.}
\end{tabular}
\end{center}
\end{table*}

\section{Concluding Remarks}\label{sec:remarks}

We summarize the points discussed in the present work.  

\begin{enumerate}

\item The isothermal X-ray temperature of a cluster, which is inferred from the X-ray surface brightness in the context of the $\beta$-model, does not always represent the virial temperature of the intracluster gas, even if the cluster is well relaxed. The virial temperature $T_{\rm vir}$ is given by $\beta T_Xx_{\rm vir}^2/(1+x_{\rm vir}^2)$ $\sim \beta T_X$ (for $x_{\rm vir} \gg 1$) rather than $T_X$ itself, where $T_X$ and $\beta$ are the isothermal gas temperature and profile-slope parameter, respectively, and $x_{\rm vir}$ is the ratio of the virial to core radii, i.e. $x_{\rm vir} = r_{\rm vir}/r_c$.  

\item For 121 clusters observed by {\it ROSAT} and {\it ASCA}, the luminosity-temperature relation $L_X - \beta T_X$, which should represent $L_X - T_{\rm vir}$, is less steep than $L_X - T_X$.  This result may imply that the $\beta$-correction to $T_X$ is important in the study of the properties expected from the self-similar collapse.

\item The 121 clusters can be classified by the core size into two groups, as suggested by \citet{OM02}.  No clear correlation is seen between $r_c$ and $r_{\rm vir}$ for the small core ($\sim$ 40--80~kpc) clusters, while the self-similarity $r_c \propto r_{\rm vir}$ is attained marginally for the larger core ($\sim$ 140--300~kpc) clusters.  Neither clear relation with cD galaxy is found for the small core clusters.  

\item It is found, however, that the small core clusters have cooling time significantly shorter than the Hubble time, while the larger core clusters have longer cooling time.  These suggest that radiative cooling in the core may be responsible for the small cores or breaking the self-similarity $r_c \propto r_{\rm vir}$, which is expected from the collapse and seen in the larger core clusters.

\item Hydrodynamical calculations of the isothermal gas in clusters, which consist of the dark matter, galaxies and gas, show the self-similar relation $r_c \propto r_{\rm vir}$ and $T_{\rm vir}\simeq \beta T$ consistently with our analysis of the $\beta$-model (section~\ref{sec:theory}).  Calculations also show that a large cD galaxy located in the cluster center affects the gas profile within the region $\sim 100$~kpc.  Thus, under the presence of a central cD galaxy, the intracluster gas may be represented better by the double $\beta$-model than the single one.  

\item However, the inner core ($\sim 40$~kpc) due to the cD galaxy is too small to account for the core-size distribution of the smaller core group.  Finally, from the analysis of the 121 clusters as well, it is unlikely that the origin of the smaller cores is explained by the presence of a central cD galaxy.

\end{enumerate}

\bigskip

The authors would like to thank Naomi Ota for her making the cluster data available and helpful suggestions for their analysis.

\end{document}